# Gender Bias in Computing

Thomas J. Misa

University of Minnesota

tmisa@umn.edu

Abstract: This paper examines the historical dimension of gender bias in the US computing workforce. It offers new quantitative data on the computing workforce prior to the availability of US Census data in the 1970s. Computer user groups (including SHARE, Inc., and the Mark IV software user group) are taken as a cross- section of the computing workforce. A novel method of gender analysis is developed to estimate women's and men's participation in computing beginning in the 1950s. The data presented here are consistent with well-known NSF statistics that show computer-science undergraduate programs enrolling increasing numbers of women students during 1965–1985. These findings challenge the "making programming masculine" thesis, and serve to correct the unrealistically high figures often cited for women's participation in early computer programming. Gender bias in computing today is traced not to 1960s professionalization but to cultural changes in the 1980s and beyond.

Gender bias in computing is fundamentally a historical problem, and it persists into the present. Computing is distinctive among all the so-called STEM fields in that computing was actually more gender-balanced three decades ago in the 1980s than it is today.  By many measures, women since the 1960s have slowly but surely gained proportional representation across the biological, physical and social sciences and the diverse engineering fields.  In most of these fields, women today hold a greater proportion of bachelor's, master's, and doctoral degrees, they form a greater proportion of faculty and researchers, and a greater share of the highly skilled technical workforce in the United States (as well as many of the technology-intensive economies of the wealthy countries of the world) than they did three or four decades ago.  For this reason, advocates of women in the sciences such as historian Margaret Rossiter can point to significant progress for women in these fields, even if obstacles remain to women's full participation in the STEM fields.[1]  Computing, however, does not fit this pattern.

---

[1] Margaret W. Rossiter, *Women Scientists in America: Forging a New World since 1972* (Johns Hopkins University Press, 2012); see perceptive reviews by Wendy Toon in *Women's History Review* 22 no. 6 (2012): 1023-1025 at doi.org/10.1080/09612025.2012.744582 and Arleen Marcia Tuchman in *American Historical Review* 118 no. 2 (2013): 538-539 at doi.org/10.1093/ahr/118.2.538

Around 1960 computing resembled the other technical fields with low representation of women in the early white-collar computing workforce and low participation in the earliest computing undergraduate degree programs. But then something unusual happened. From 1965 to 1985, women gained an increasing proportion of undergraduate computer science degrees, one readily accessible if obviously incomplete measure of the emerging field, fully tripling across these years from around 12 percent to 37 percent. These twenty years witnessed great intellectual and institutional changes in the field of computer science, and great expansion, but all the same no other technical field in the post-1945 era of higher education has experienced such swift growth in women's participation. Similarly, although the national statistics are incomplete, women experienced significant growth in participation and absolute numbers in the white-collar computing workforce. U.S. Department of Labor statistics compiled for the Standard Industrial Classification (SIC) indicate that women's participation in the computer *manufacturing* workforce increased from 27 to 31 percent during 1967–1974; and by the mid-1980s women's participation in the white-collar computing workforce had risen all the way to 38 percent.[2] These impressive numbers were a powerful positive corrective to those in later decades who floated ill-advised suggestions that somehow women didn't like computing or even, as Harvard's Larry Summers infamously put it, that "issues of intrinsic aptitude" made women ill-suited to careers in technical fields.[3] Obviously, since computing was at a certain moment nearly half women, these half-baked suggestions mostly fell flat.

Then in the mid-1980s came the second historically distinctive development in computing. Women's proportion of computer science undergraduate degrees in the U.S. peaked —and then started falling dramatically—with the numbers going down to around 15 or 20 percent by the early 2000s, depending on which statistics are consulted, and with women's absolute numbers falling steeply. Computer science degrees awarded to women during 1985–1995 fell by more than half from 14,431 to 7,063 while those awarded to men dropped around a quarter from 24,690 to 17,706. Generally, women's share of master's degrees in computer science peaked in the mid-1980s at 30 percent and then held steady for 15 years, while women's share of doctoral degrees experienced slow if unsteady growth throughout the 1960s–2000s. In

---

[2] Richard E. Weber and Bruce Gilchrist, "Discrimination in the Employment of Women in the Computer Industry," *Communications of the ACM* 18 no. 7 (1975): 416-418 at dx.doi.org/10.1145/360881.360921 . For historical statistics, see Caroline Clarke Hayes, "Computer Science: The Incredible Shrinking Woman," in Thomas J. Misa, ed., *Gender Codes: Why Women Are Leaving Computing* (John Wiley, 2010), 25-49. A valuable analysis of the IT workforce since 1970—including assessment of the evolving Census categories used to analyze it—is Julia Beckhusen,"Occupations in Information Technology," *American Community Survey Reports* ACS-35 (Washington DC: U.S. Census Bureau, 2016) at www.census.gov/content/dam/Census/library/publications/2016/acs/acs-35.pdf. One mistake in this report, a consequence of its following the *decade-by-decade* Census data is that it does not pick up the *1985* peak, instead asserting (using data from 1970, 1980, 1990 et seq) that "The percentage of women working in IT occupations peaked in *1990* [emphasis added] at 31.0 percent."

[3] Scott Jaschik, "What Larry Summers Said," *Inside Higher Ed* (18 February 2005) at www.insidehighered.com/news/2005/02/18/summers2_18

Rossiter's words, for undergraduate women there was a "collapse … in computer science … after 1985."[4]  Computer science became something of a boy's club.  Generally, during these years, the most prestigious computer science departments experienced precipitous drops in their enrollment of women,[5] but the trend existed across virtually all U.S. computer science programs; and it has persisted so that today the proportion of women gaining undergraduate computer science degrees, apart from a few notable success stories, is near where it was in the 1960s.[6] Many of the OECD countries followed these U.S. trends.[7]

This collapse in women's undergraduate enrollments in computing—computer science, information science, and similar computing-centered degree programs—has attracted a great deal of attention by the computing profession, the educational world, and policy actors.[8]  And the problem is not at all confined to the U.S.  For 21 countries based on OECD data from 2001, researchers found substantial "male over-representation" across the board in undergraduate computing-degree programs ranging from a low of 1.79 in Turkey to a high of 6.42 in the Czech Republic; the United States was a middling 2.10, with these figures corrected for the underlying male:female enrollments in each country's higher-education system.[9]  It is modestly good news that women have not been further left behind with the current boom in computer science, as total undergraduate computer science majors are recently up by 300 percent (2006-2015). Still, as a recent analysis reminds us, "as previous enrollment surges [in the mid-1980s and early 2000s] waned, interest in computing by females dropped more significantly than for males and has never recovered to previous levels."[10]

---

[4] Margaret W. Rossiter, *Women Scientists in America: Forging a New World since 1972* (Johns Hopkins University Press, 2012), quote p. 41 (collapse).  Compare Rossiter's graphs for computer science (figure 3.11) with other fields (figures 3.5 to 3.10). For computer science degrees, see Susan T. Hill, "Science and Engineering Bachelor's Degrees Awarded to Women Increase Overall, but Decline in Several Fields," NSF 97-326 (7 November 1997) at www.nsf.gov/statistics/databrf/sdb97326.htm

[5] According to the well-respected CRA Taulbee survey of doctoral-granting departments, the low point in women's share of undergraduate computer science degrees was 11.2 percent in 2009.  See data at www.cra.org/resources/crn/, www.cra.org/resources/taulbee, and ncsesdata.nsf.gov/webcaspar/ .

[6] Carnegie Mellon, Harvey Mudd, and University of California–Berkeley are widely discussed recent success stories for women in undergraduate computer science.  See Sarah McBride, "Glimmers of Hope for Women in the Male-Dominated Tech Industry," Bloomberg Technology (8 March 2018) at www.bloomberg.com/news/articles/2018-03-08/glimmers-of-hope-for-women-in-the-male-dominated-tech-industry

[7] "An analysis of computer science shows a steady decrease in female graduates since 2000 that is particularly marked in high-income countries," reports UNESCO in "Women still a minority in engineering and computer science" (2015) at http://www.unesco.org/new/en/media-services/single-view/news/women_still_a_minority_in_engineering_and_computer_science/. See also Vashti Galpin, "Women in computing around the world," *SIGCSE Bulletin* 34 no. 2 (June 2002): 94-100. doi.org/10.1145/543812.543839 ; Merete Lie, "Technology and Masculinity: The Case of the Computer," *European Journal of Women's Studies* 2 no. 3 (1995): 379-394 at doi.org/10.1177%2F135050689500200306; *UNESCO Science Report: Towards 2030* (UNESCO Publishing, 2015) at unesdoc.unesco.org/images/0023/002354/235406e.pdf .

[8] It is essential to acknowledge that academic computer science is only one route, among many, to the computing workforce.  Indeed, "most IT workers receive their formal education in fields other than computer science," according to Peter Freeman and William Aspray, *The Supply of Information Technology Workers in the United States* (Computing Research Association, 1999), quote p. 17 at archive.cra.org/reports/wits/it_worker_shortage_book.pdf. The authors list no fewer than 20 "IT-related Academic Disciplines Offered in the United States" (table 2-1 on p. 28).  Diverse computing disciplines—such as software engineering, computer engineering, computational science, information systems, information science, and others, in addition to computer science—contribute to the computing profession, in the view of Peter Denning, "Computing the Profession," *Educom Review* 33 no. 6 (1998): 26-30, 46-59 at www.educause.edu/ir/library/html/erm/erm98/erm9862.html

[9] Maria Charles and Karen Bradley, "A Matter of Degrees: Female Underrepresentation in Computer Science Programs Cross-Nationally," in J. McGrath Cohoon and William Aspray, eds., *Women and Information Technology: Research on Underrepresentation* (MIT Press, 2006),  183-203 on 190.

Even worse for the wider economy, the proportion of women in the skilled computing workforce in the United States also began *dropping* in the late 1980s, clearly indicating that the problem was not merely one in academic computer science. In the 2011 American Community Survey from the U.S. Census, women constituted just 27 percent of the computing workforce, down more than 10 percentage points from the mid-1980s peak—a decline by more than one fourth.[11] And despite composing 48 percent of the entire U.S. workforce, women represent around half that share in the computing workforce; and since the computing workforce now accounts for fully 50 percent of the STEM workforce, women's underrepresentation in computing has wide ramifications.[12] In recent years, an avalanche of journalism has lamented the low participation of women in the tech workforce and documented the persistence of harrowing and offensive sexism.[13] Women are on the margins of technical jobs at top Silicon Vally companies, ranging, according to 2015 figures, from Apple (20 percent women), through Google and LinkedIn (both 17) and Facebook and Yahoo (both 15), down to Twitter (10 percent).[14]  Men outnumber women 10:1 in Silicon Valley's executive positions and 40:1 in volume of venture-capital funding.[15]  Uber's CEO Travis Kalanick became a demented poster child for endemic tech sexism, leading to his ouster in June 2017.[16]  And even at image-conscious Google there was the attention-grabbing internal memo asserting "the distribution of preferences and abilities of men and women differ in part due to biological causes" which (it was claimed) leads to women's low participation in tech jobs and tech leadership.[17]

---

[10] Eric Roberts, Tracy Camp, David Culler, Charles Isbell, and Jodi Tims, "Rising CS Enrollments: Meeting the Challenges," in *Proceedings of the 49th ACM Technical Symposium on Computer Science Education* (New York: ACM, 2018), 539-540 at doi.org/10.1145/3159450.3159628.

[11] One can acknowledge increases in the absolute numbers of women, since expansion in the IT workforce offsets declines in female participation. The U.S. IT workforce was 781,000 in 1980, 1.5 million in 1990, 3.4 million in 2000, and 4.0 million in 2010, according to Julia Beckhusen,"Occupations in Information Technology," *American Community Survey Reports* ACS-35 (Washington DC: U.S. Census Bureau, 2016), p. 2.

[12] Liana Christin Landivar, "Disparities in STEM Employment by Sex, Race, and Hispanic Origin," U.S. Census American Community Survey Reports (September 2013) at www.census.gov/prod/2013pubs/acs-24.pdf ; on pp. 4, 6. With greater detail, the ACS table 3 reports women at 26.6 percent of the computing workforce, ranging across 12 sub-categories from a high of 40.1% of database administrators to a low of 11.4% of computer network architects. The largest sub-category is software developers, comprising a full 11.8% of the entire STEM workforce, with 22.1% women. The AAUW's analysis of Census data reported women computer professionals in 11 sub-categories ranging from a high of 39% for web developers to a low of 7% for network architects (with database administrators at 32% women); see Christianne Corbett and Catherine Hill, *Solving the Equation: The Variables for Women's Success in Engineering and Computing* (Washington DC: AAUW, 2015), p. 14 at www.aauw.org/research/solving-the-equation/.

[13] Jonny Evans, "The technology world's sexism needs to end," *Computerworld* (30 April 2014) at www.computerworld.com/article/2488845/it-management/the-technology-world-s-sexism-needs-to-end.html ; Zachary Jason, "Game of Fear" [on so-called Gamergate], *Boston Magazine* (28 April 2015) at www.bostonmagazine.com/news/2015/04/28/gamergate/ ; Liza Mundy, "Why Is Silicon Valley So Awful to Women?" *Atlantic Monthly* (April 2017) at www.theatlantic.com/magazine/archive/2017/04/why-is-silicon-valley-so-awful-to-women/517788/ ; Katie Benner, "Women in Tech Speak Frankly on Culture of Harassment," *New York Times* (30 June 2017) at www.nytimes.com/2017/06/30/technology/women-entrepreneurs-speak-out-sexual-harassment.html ; Sheelah Kolhatkar, "The Tech Industry's Gender-Discrimination Problem," *New Yorker* (20 November 2017) at www.newyorker.com/magazine/2017/11/20/the-tech-industrys-gender-discrimination-problem ; and Emily Chang, *Brotopia: Breaking Up the Boys' Club of Silicon Valley* (New York: Portfolio, 2018).

As can readily be imagined, the magnitude of gender bias in computing has generated an immense and dauntingly diverse literature.  There is alas no easy answer to the question "what caused" the dramatic fall in women's participation in computing, and a great many have offered suggestions about "what is to be done?"  Policy actors such as the National Science Foundation, the National Center for Women in Information Technology, the Anita Borg Institute for Women and Technology and its now-annual Grace Hopper Celebration of Women in Computing, the Alfred P. Sloan Foundation, as well as professional groups, such as the Computing Research Association's Committee on the Status of Women in Computing Research (CRA–W) and the Association for Computing Machinery's Committee on Women in Computing (ACM–W), have debated, proposed, and enacted numerous initiatives to correct women's underrepresentation.[18]  These include attention to systemic issues in the computing curriculum, classroom culture, recruitment, and retention as well as more focused interventions such as peer programming.  For their 2006 edited volume, *Women and Information Technology: Research on Underrepresentation*, Joanne McGrath Cohoon and William Aspray surveyed the voluminous social-science literature and came to the sobering conclusion that "twenty-five years of interventions have not worked."[19]  Recently Aspray published two Sloan-supported volumes narrating NSF's efforts at broadening participation in computing and describing the experiences of women, African-Americans, Hispanics, and Native Americans in the field.[20]  While much of the literature focuses on the United States, there are suggestive case studies from around the world and three book-length treatments that pay sustained attention to Europe.[21]

---

[14] Dave Smith, "Women Are Vastly Underrepresented In The Tech Sector," *Business Insider* (14 August 2014) at www.businessinsider.com/chart-of-the-day-women-are-vastly-underrepresented-in-the-tech-sector-2014-8 ; for recent figures see Will Evans and Sinduja Rangarajan, "Hidden figures: How Silicon Valley keeps diversity data secret," *Reveal* (19 October 2017) at www.revealnews.org/article/hidden-figures-how-silicon-valley-keeps-diversity-data-secret/

[15] Maya Kosoff, "Here's Evidence That It's Still Not A Great Time To Be A Woman In Silicon Valley," *Business Insider* (2 January 2015) at www.businessinsider.com/women-hold-just-11-of-executive-positions-at-silicon-valley-tech-companies-2015-1 ; Valentina Zarya, "Venture Capital's Funding Gender Gap Is Actually Getting Worse," *Fortune* (13 March 2017) at fortune.com/2017/03/13/female-founders-venture-capital/

[16] Susan Fowler, "Reflecting on one very, very strange year at Uber," (19 February 2017) at www.susanjfowler.com/blog/2017/2/19/reflecting-on-one-very-strange-year-at-uber ; Mike Isaac, "Inside Travis Kalanick's Resignation as Uber's C.E.O." *New York Times* (21 June 2017) at www.nytimes.com/2017/06/21/technology/uber-travis-kalanick-final-hours.html

[17] Kate Conger, "Here's The Full 10-Page Anti-Diversity Screed Circulating Internally at Google," *Gizmodo* (5 August 2017) at gizmodo.com/exclusive-heres-the-full-10-page-anti-diversity-screed-1797564320 ; Rosalind C. Barnett and Caryl Rivers, "We've studied gender and STEM for 25 years: The science doesn't support the Google memo," *Recode* (11 August 2017) at www.recode.net/2017/8/11/16127992/google-engineer-memo-research-science-women-biology-tech-james-damore

[18] Amy Sue Bix, "Organized Advocacy for Professional Women in Computing: Comparing Histories of the AWC and ACM–W," in Thomas J. Misa, ed., *Communities of Computing: Computer Science and Society in the ACM* (New York: Association for Computing Machinery and Morgan & Claypool, 2016), 143-172 at doi.org/10.1145/2973856.2973864 .

Naturally, academic historians of computing have engaged the problem of gender bias.  Historians Jennifer Light, Nathan Ensmenger, Janet Abbate, and Marie Hicks have each contributed to raising the visibility of women in early computing.  The suggestion is even that early computer programming was dominated by women.  In her well-cited *Technology and Culture* article, "When Computers Were Women," Light points to an idiom of sex-typing that was pervasive during and after the Second World War: "designing [computer] hardware was a man's job; programming was a woman's job"—and goes on to describe "how the job of programmer, perceived in recent years as masculine work, originated as feminized clerical labor."[22]  Women such as Grace Hopper, Jean Jennings, Frances Elizabeth Holberton, and dozens of others certainly were prominent in early computer programming.  "The exact percentage of female programmers is difficult to pin down with any accuracy," writes Ensmenger in *Gender Codes*, "but … reliable contemporary observers suggest that it was [close] to 30 percent."  Elsewhere he suggests women were as much as *50* percent of computer programmers in the years before male-dominated professionalization and garden-variety sexism resulted in pushing them aside and "making programming masculine."[23]  In a follow-on article Ensmenger points to "the masculinization of computer programming" during the 1960s and early 1970s (note the years) that resulted in the distinctive, pervasive, and permanent masculine culture in computing.[24] In her recent prize-winning book *Programming Inequality*, Hicks widens these observations to suggest that Britain lost its early lead in computing (its proto-computers for breaking the German wartime Enigma and Lorenz ciphers, although shrouded in secrecy, were foundational for the first stored-program digital computers at Manchester and Cambridge universities) because the country shunted its largely female computing workforce into dead-end jobs. Hicks specifically includes both highly skilled programmers and analysts as well as lower-skilled operators and technicians, reminding us that women up and down the status hierarchy made contributions to getting early computers to do useful work. Focusing more on "the upper echelon of the computing field," Janet Abbate's recent *Recoding Gender* is based on 52 interviews with eminent professional women in the US and UK with the aim "to make visible some notable contributions by women."[25]

---

[19] MIT Press (2006), quote p. ix.

[20] William Aspray, *Participation in Computing: The National Science Foundation's Expansionary Programs* (Springer International, 2016) at doi.org/10.1007/978-3-319-24832-5 ; and idem, *Women and Underrepresented Minorities in Computing: A Historical and Social Study* (Springer International, 2016) at doi.org/10.1007/978-3-319-24811-0 .

[21] Merete Lie, ed., *He, She and IT Revisited: New Perspectives on Gender in the Information Society* (Oslo: Gyldendal Akademisk, 2003); Thomas J. Misa, ed., *Gender Codes: Why Women Are Leaving Computing* (John Wiley, 2010); and Valérie Schafer and Benjamin G. Thierry, eds., *Connecting Women: Women, Gender and ICT in Europe in the Nineteenth and Twentieth Century* (Springer International, 2015).  Influential international studies include Vivian Anette Lagesen, "A Cyberfeminist Utopia?: Perceptions of Gender and Computer Science among Malaysian Women Computer Science Students and Faculty," *Science, Technology & Human Values* 33 no. 1 (2008): 5-27 at doi.org/10.1177/0162243907306192 ; Ulf Mellström, "The Intersection of Gender, Race and Cultural Boundaries, or Why is Computer Science in Malaysia Dominated by Women?," *Social Studies of Science* 39 no. 6 (2009): 885-907 at doi.org/10.1177/0306312709334636 ; and Roli Varma and Deepak Kapur, "Decoding Femininity in Computer Science in India," *Communications of the ACM* 58 no. 5 (2015): 56-62 doi.org/10.1145/2663339

[22] Jennifer S. Light, "When Computers Were Women," *Technology and Culture* 40 no. 3 (1999): 455-483, quotes on 455 (job of programmer) and 469 (man's job) at www.jstor.org/stable/25147356. An earlier article documenting this history was W. Barkley Fritz, "The Women of ENIAC," *IEEE Annals of the History of Computing* 18 no. 3 (1996): 13-28 at doi.org/10.1109/85.511940

[23] Nathan Ensmenger, "Making Programming Masculine," in Thomas J. Misa, ed., *Gender Codes: Why Women Are Leaving Computing* (John Wiley, 2010), 115-141, quote p. 116.  For the claim of 50 percent, see the unedited Ensmenger chapter at homes.soic.indiana.edu/nensmeng/files/ensmenger-gender.pdf (accessed January 2018) on p. 2.

It is fascinating to watch the transformation of a historian's conjecture into the certainty of a widely circulated "meme" broadcast to the public by the Smithsonian, National Public Radio, and the *Wall Street Journal*.[26] It seems the conventional wisdom now is that while men dominated the hardware side, "computer programming was a women's field" and that "computer programming was a feminized occupation from its origins."[27] Historians' nuanced discussion of women in early computing was popularized by Walter Isaacson in his best-selling *The Innovators* (2014) and subsequently amplified by journalists, bloggers, and film makers.[28] Along the way, the numbers of women grew ever more impressive. "Between 30 and 50 percent of programmers were women in the 1950s," according to one oft-repeated meme.[29] It seemed (in another repeated meme) that "men's takeover of the field in the late 1960s [led to] an immense climb in pay and prestige."[30] "The decline in female programmers coincided with the professionalization of coding in the 1960s," writes the *Wall Street Journal*.[31]

---

[24] Nathan Ensmenger, "'Beards, Sandals, and Other Signs of Rugged Individualism': Masculine Culture within the Computing Professions," *Osiris* 30 (2015): 38-65 at doi.org/10.1086/682955

[25] Marie Hicks, *Programmed Inequality: How Britain Discarded Women Technologists and Lost Its Edge in Computing* (MIT Press, 2017); Janet Abbate, *Recoding Gender: Women's Changing Participation in Computing* (MIT Press, 2012), quote p. 7. Corinna Schlombs explores the wider sense of "gender" not limited to women's history *per se* in her "The 'IBM Family': American Welfare Capitalism, Labor, and Gender in Postwar Germany," *IEEE Annals of the History of Computing* 39 no. 4 (2017): 12-26 at doi.org/10.1109/MAHC.2018.1221046 . Like Hicks, Thomas Haigh includes both higher and lower skilled women in his analysis of the data processing workforce; see Thomas Haigh, "Masculinity and the Machine Man: Gender in the History of Data Processing," in Misa, Gender Codes, 51-71. By comparison, my concerns are the higher skilled or white collar computing (or information technology) workforce. In 1970, the Census used three subcategories (computer programmers, computer systems analysts, and "all other" computer specialists) and by 2010 it used 12 sub-categories; see Julia Beckhusen, "Occupations in Information Technology," *American Community Survey Reports* ACS-35 (Washington DC: U.S. Census Bureau, 2016), 3-6.

[26] The claim of computer programming being, at any time, 50 percent women is thinly sourced. Ensmenger's source for the "reliable contemporary observers" claiming 30 to 50 percent women is Richard Canning, "Issues in Programming Management," *EDP Analyzer* 12 no. 4 (1974): 1-14. It is also the source—besides an incompletely cited article in the trade journal *Datamation* (1964) that is mis-attributed to sociologist Sherry Turkle—supporting his later claim (2015: quote p. 59) "in most corporations women represented at least 25–30 percent of all computer personnel" specifically not including the highly feminized computer and keypunch operators which, if they were included, "the representation of women would be even higher." Women are mentioned on two pages of the 1974 Canning article: a manager with IBM Federal Systems Division stated that, for one IBM programming group, "about one-half the programmers are women, and … the number of women managers is rising rapidly" (p. 2); and in a different context "a woman team member might in fact play the moderating role of 'mother'." (p. 5). Canning's quote that "the number of women managers is rising rapidly" is consistent with women entering the computing workforce in the 1970s and is obviously inconsistent with the counterfactual assertion that women were leaving computing in the 1970s.

I think the process that connects an academic conjecture to the certainty of internet memes goes something like this. We ache for some comprehensible understanding to the origin of gender bias in computing. The notion that computer programming was born female and then made masculine, and that this history has passed straight down to the present day, seems plausible. It has the great attraction of a linear storyline or plot: the world was once some way (women dominated computer programming), then it changed (programming was made masculine), and that led directly to the present moment, where quite obviously men dominate computing. Ensmenger's claim of 30 or even 50 percent women in computer programming, launched in academic publications and available on the world wide web, gained a wide audience through his interview for a popular film "Code: Debugging the Gender Gap" (2015) done by

---

[27] Quotes, respectively, from Rose Eveleth, "Computer Programming Used To Be Women's Work" smithsonian.com (7 October 2013) at www.smithsonianmag.com/smart-news/computer-programming-used-to-be-womens-work-718061/ and Ensmenger, "Beards, Sandals, and Other Signs of Rugged Individualism," p. 44. "Decades ago, it was women who pioneered computer programming," according to Laura Sydell, "The Forgotten Female Programmers Who Created Modern Tech," NPR Morning Edition (6 October 2014) at www.npr.org/sections/alltechconsidered/2014/10/06/345799830/the-forgotten-female-programmers-who-created-modern-tech

[28] For a critical review, see Thomas Haigh and Mark Priestley, "Innovators Assemble: Ada Lovelace, Walter Isaacson, and the Superheroines of Computing," *Communications of the ACM* 58 no. 9 (2015): 20-27 at doi.org/10.1145/2804228

[29] Josh O'Connor, "Women pioneered computer programming: Then men took their industry over," Timeline (16 May 2017) at timeline.com/women-pioneered-computer-programming-then-men-took-their-industry-over-c2959b822523 . See numerous additional citations in note 33.

[30] "What Programming's Past Reveals About Today's Gender-Pay Gap," *Atlantic Monthly* (September 2016) at www.theatlantic.com/business/archive/2016/09/what-programmings-past-reveals-about-todays-gender-pay-gap/498797/ . "In the 1950s and '60s, employers began relying on aptitude tests and personality profiles that weeded out women by prioritizing stereotypically masculine traits and, increasingly, antisocialness," according to Becky Little, "The First 1940s Coders Were Women—So How Did Tech Bros Take Over?" *History Channel* (1 September 2017) at www.history.com/news/coding-used-to-be-a-womans-job-so-it-was-paid-less-and-undervalued

[31] Christopher Mims, "The First Women in Tech Didn't Leave—Men Pushed Them Out," *Wall Street Journal* (10 December 2017) at www.wsj.com/articles/the-first-women-in-tech-didnt-leavemen-pushed-them-out-1512907200. In three paragraphs the logical inconsistency is revealed: "The decline in female programmers coincided with the professionalization of coding in the 1960s, writes computer historian Nathan Ensmenger in his 2012 book *The Computer Boys Take Over*. . . . The proportion of women earning degrees in computer science peaked in 1984 at 37%." (emphasis added)

Robin Hauser Reynolds.[32]  This film then became the source for numerous confident assertions that "women made up 30 percent to 50 percent of all programmers."[33]

Only *one* of the three above widely publicized "memes" about women in early computing is plausibly true.  Computer programming was a booming and lucrative field in the 1960s.  The other claims are not well grounded.  The commonly held view of computing women during these early decades leaves a lot to desire.  Let's consider each of these assertions—before presenting this chapter's new data that corrects our understanding.  Getting the history correct—when did women leave computing?—is essential to correctly perceiving the current problem of gender bias in computing.

First, while women were clearly prominent in early computing and played critical roles in developing computer programming, it is inaccurate to claim that women composed half the professional or highly skilled members of the early field.  Ground zero for our understanding of women in computing has been the "women of ENIAC," Grace Hopper, and their many women colleagues' remarkable achievements and unusual prominence.  In 1949 at an international computing conference at Harvard University there were 33 notable women that form a who's who for women in computing, with high-level representation from Harvard, MIT, Raytheon, the U.S. National Bureau of Standards, Census Bureau, and three military agencies, among other computing hotspots at the time.  Mina Rees from the Office of Naval Research chaired a three-hour session on "Recent Developments in Computing Machinery" with heavyweight contributions from Bell Telephone Laboratories, General Electric, Raytheon, Eckert-Mauchly Computer Company, Harvard, and MIT; but she was the *only woman* on the four-day program.  In addition to the 33 female attendees, there were 540 male attendees who can be identified, and so women comprised around 6 percent of the Harvard conference.  This chapter analyzes new data from the 1950s through 1980s and estimates that women were roughly 15 percent of the computing field (see photograph) as it developed into a highly skilled and highly paid profession.

---

[32] See "Code: Debugging the Gender Gap," (2015) at https://www.codedoc.co/ and Stephen Cass, "A Review of *Code: Debugging the Gender Gap*," *IEEE Spectrum* (19 June 2015) at spectrum.ieee.org/geek-life/reviews/a-review-of-code-debugging-the-gender-gap

[33] "By the 1960s, women made up 30% to 50% of all programmers, according to Ensmenger" (specifically citing the film), states Jane Porter, "The Fascinating Evolution of Brogramming And The Fight To Get Women Back," *Fast Company* (20 October 2014) at https://www.fastcompany.com/3037269/the-fascinating-evolution-of-brogramming-and-the-fight-to-get-women-back.  "50 years ago, half of computer programmers were women," affording to Lulu Chang, "Indiegogo Documentary 'Code' Asks Why Women Stopped Coding, & It's An Important Question," *Bustle* (26 October 2014) at www.bustle.com/articles/45960-indiegogo-documentary-code-asks-why-women-stopped-coding-its-an-important-question .  "Between 30 and 50 percent of programmers were women in the 1950s" according to Palak Kapadia, "Computer Programming Started Off as Woman's Job," *Sheroes* (29 May 2017) at sheroes.com/articles/did-you-know-computer-programming-started-off-as-women-s-job/NTI2MQ== .  "Between 30 and 50 percent of programmers were women in the 1950s," repeats Rebel Girls on Facebook (8 June 2017) at www.facebook.com/rebelgirls/posts/1580025575364635 .  "In the 1950's, 30 to 50 percent of computer programmers were women," reiterates Chava Shapiro in "Women Used To Dominate Tech… Until They Were Pushed Out," *The Wisdom Daily* (10 August 2017) at thewisdomdaily.com/women-used-to-dominate-tech-until-men-pushed-them-out/ .

Second, the oft-repeated suggestion that men staged a take-over of computing sometime in the 1960s and pushed women aside is simply wrong. Women *did not leave* the computing field to men in the 1950s or 1960s or 1970s; quite the opposite. As noted above, women gained an *increasing* proportion of computer science bachelor's degrees between 1965 and 1985, and women formed an increasing proportion of the white-collar computing workforce through the 1980s. During the very years when the entrenched popular meme has it (incorrectly) that men were chasing women out of computing, women were actually flooding into computing.

Third, I believe that getting the history correct is necessary to properly understand gender bias in computing and the tech industry. In the 1950s and 1960s women, notwithstanding their achievements in computing, were soundly outnumbered by men, as data in this chapter will demonstrate. Through the 1960s and 1970s women's participation in computing was steadily *increasing*. Only in the 1980s did women's participation in computing began shrinking and, from then, lead to today's situation. We cannot understand present-day gender bias in computing as a product of 1960s sexism but rather need to understand the later developments of the 1980s.[34]

This chapter supports these three observations with newly collected data from the 1950s through 1980s. It first introduces a method developed at CBI to extract meaningful and systematic data on women in computing before 1970. It then discusses two prominent computer user groups whose records permit coverage of the years 1955 to 1989. This chapter is drawn from a larger book-length study on women in the computing industry.

**New data on computing women before 1970**

All data on large populations depends on statistical methods and proper sampling. For the 1970 U.S. Census a 20 percent sample of U.S. households were asked about their occupations, and from this sample comes the figure of 22.5 percent women in the U.S. computing workforce, widely cited as the first reliable figure.[35] Earlier censuses did not separately tabulate women in the computing workforce. I do not claim that my three data samples reported below, individually, are perfect. Nevertheless, as we shall shortly see, these varied samples do have the

---

[34] In "Making Programming Masculine," his *Gender Codes* chapter, Ensmenger cites instances of egregious sexism sourced from the trade journal *Datamation* (13 citations) from the 1960s. But he overlooks the changes in the 1970s and 1980s in computing's gender composition and the changed cultural climate in the computing industry and profession.

[35] Bruce Gilchrist and Richard E. Weber, "Enumerating full-time Programmers," *Communications of the ACM* 17 no. 10 (October 1974): 592-593 at dx.doi.org/10.1145/355620.361177 . In turn, the Census figure of 22.5 percent women is consistent with a 600,000-person salary survey done in 1971 by *Business Automation* which found "women made up 14% of systems analysts and 21% of computer programmers," according to Thomas Haigh in "Masculinity and the Machine Man: Gender in the History of Data Processing," in Misa, *Gender Codes*, 51-71, quote p. 64.

**Figure 1: Harvard Mark 1 team in 1945**

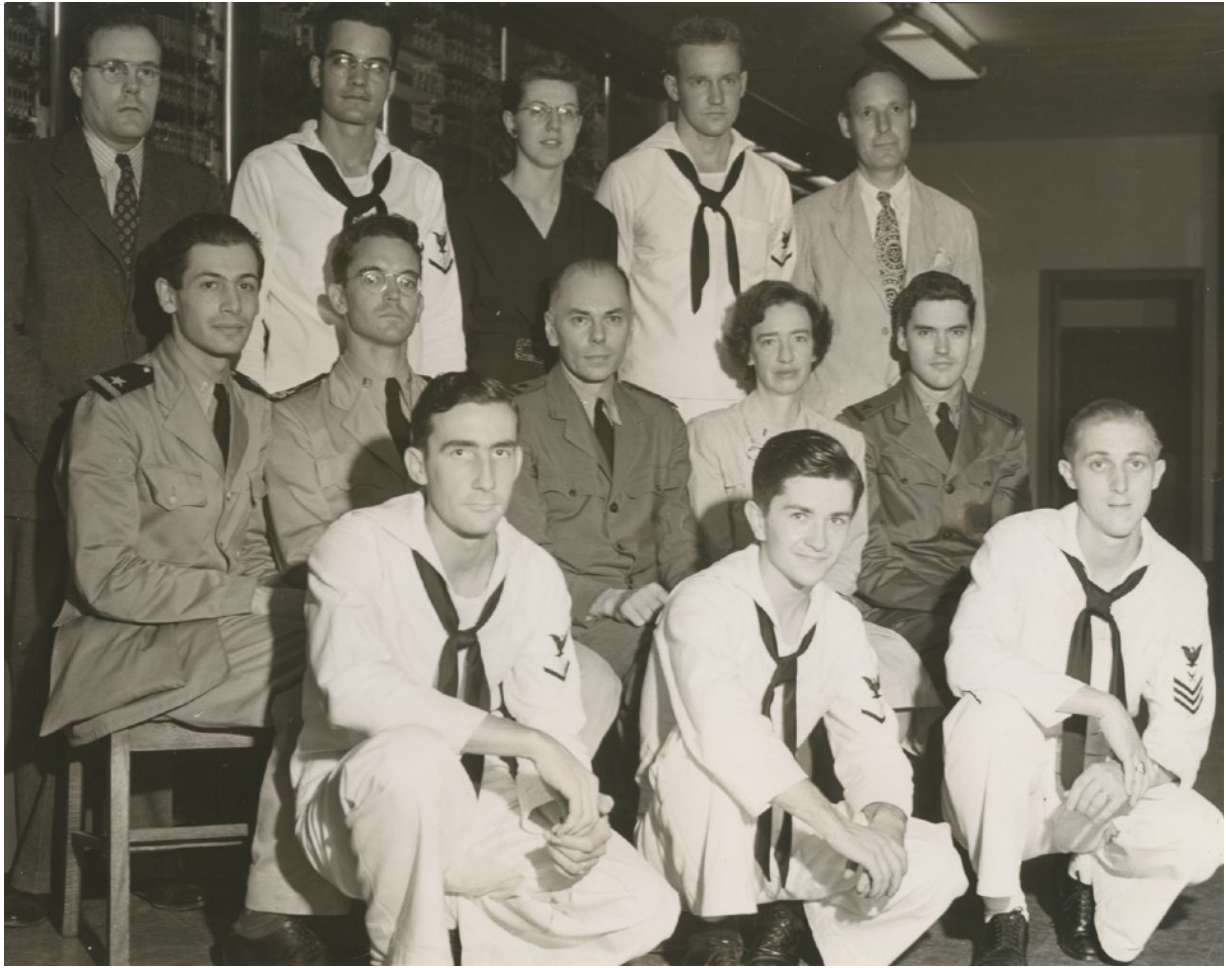

An image of computing as 15 percent women (2 women in 13) with Lieutenant Grace Hopper in second row and computer operator Ruth Knowlton behind Commander Howard Aiken.

virtue of consistency: they indicate a reliable pattern that suggests computing women were around 5 percent of the field in the 1950s and then increased to roughly 15 percent by the 1960s (see figure 1) and continued rising into the 1980s.  I next discuss the research method used to create this new data series.

     Computing conferences, professional societies, and user groups prepared attendee and member lists that are now available in their organizational archives.  Many of these lists included both first (given) and last (family) names.  On a suggestion by CBI's Jeffrey Yost, and after refinement by William Vogel, I examined CBI's set of user-group archival records with a sizable

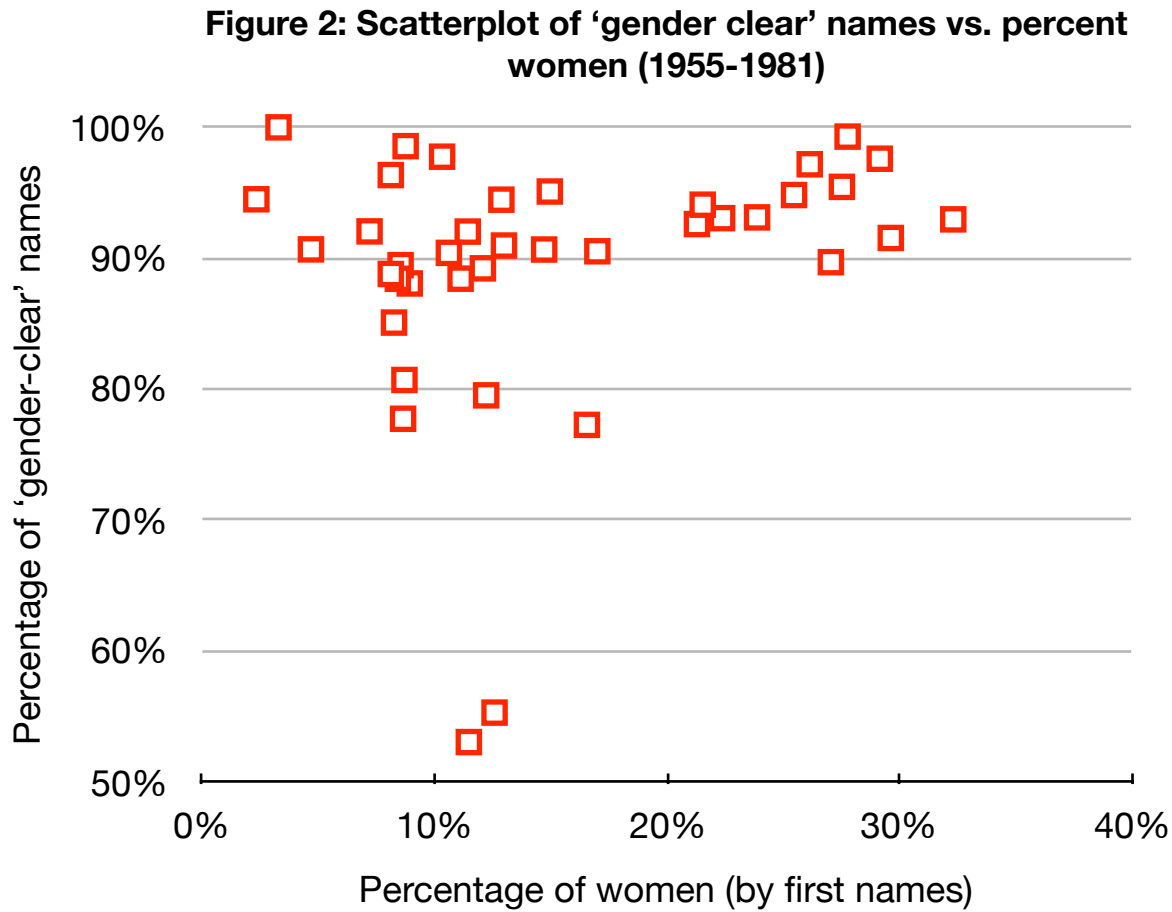

sample reported here.36 It's a simple matter to count the number of Margaret, Betty, Mary, and Dorothy's in these lists and tally against the number of William, George, Robert, and Edward's. To resolve cases of initials-only attendees, one can look for the gender-revealing first names often given in accompanying documentation; explicit references to "Mr" or "Mrs" or "Miss" can resolve gender-ambiguous first names and initials. What is more, the Social Security Administration published thousands of the most-common first names—ranked by frequency of their use and identified by gender—year-by-year beginning in 1880.37 Names change. Whereas "Robin" was a gender-ambiguous name for people born in 1930 (7:5 male) and becomes a woman's name by 1960 (10:1 female), "Leslie" actually changes gender between 1930 (9:1 male) and 1960 (3:1 female).  The short name "Pat" remains gender-ambiguous throughout.  In this way, instances of most U.S. names can be resolved with historical accuracy.[38]  Persons with initials-only or gender-ambiguous names were sometimes resolved by 'linking' the specific person to gender-clear identifications in other meetings or publications or oral histories.  Overall, as the scatterplot indicates (see figure 2), typically 80 to 100 percent of individuals in this data set can be clearly identified by gender, even as the percentage of women varied from around 3 to just over 30 percent.  The two low-ball figures (just over 50 percent gender-clear names) are discussed below.

---

36 William F. Vogel, "'The Spitting Image of a Woman Programmer': Changing Portrayals of Women in the American Computing Industry, 1958-1985," *IEEE Annals of the History of Computing* 39 no. 2 (2017): 49-64. I have done preliminary analysis also of data from early computer conferences and membership lists (1948-1955) and two other user groups.

37 See data at www.ssa.gov/OACT/babynames/limits.html. The dataset is elsewhere described as "a 100 percent sample of Social Security card applications after 1879" (trimmed to suppress first names with fewer than 5 instances); see https://catalog.data.gov/dataset/baby-names-from-social-security-card-applications-national-level-data

For each computer user-group list, I computed the percentage of women in the (gender-identified) total of men and women; and have plotted as time series these percentages along with the total size of the user group meeting in the graphs below.  For the percentages, both the numerators and denominators set aside the gender-ambiguous "Pat's" and initials-only attendees, if they could not be resolved, while the total size includes all attendees for each meeting.  This data on women in early professional computing gives insight into computing's gender balance in the decades before the government statistics are available.

**Data from IBM user group SHARE**

The computer industry's prominent user groups began in 1955 with the organizational users of IBM and UNIVAC computers, initially centered in southern California's aerospace industry, with major computing efforts at Ramo-Woodridge, Douglas Aircraft, Hughes Aircraft, Lockheed Aircraft, North American Aviation, and RAND.  The founding meetings of SHARE and USE, both in 1955, were held at RAND and Ramo-Woodridge, respectively; and both user groups quickly attracted nation-wide participation.  These included government facilities at Los Alamos, Livermore, the National Security Agency, and the Census Bureau; Boeing Airplane in Seattle; corporations such as General Electric and General Motors; east coast aviation companies Curtiss-Wright and United Aircraft (a spin-off from Boeing); and other users of these large-scale machines.  Since SHARE meetings included representatives of computer users and the computer manufacturer IBM, the user group data sheds light on both computer users' and manufacturers' employment of women.  By the early 1970s, nearly two thousand people attended SHARE's twice-yearly meetings. The IBM user group SHARE has been profiled in articles by Atsushi Akera and by Jeffrey Yost.[39]

---

[38] I verified this method with a list of 228 women who gained PhD's in math before 1940, scoring 223 correctly as female, 0 incorrectly as male, and 5 or 6 gender-ambiguous names (Wealthy, Shu Ting, Abba, Andrewa, Echo, Bird). SSA's 1900 year-of-birth data does not list these 5 names nor "Bird"; its 1880 data identifies "Bird" as female. See Judy Green and Jeanne LaDuke, *Pioneering Women in American Mathematics: The Pre‒1940 Ph.D.'s* (Providence RI: American Mathematical Society, 2009 / London: London Mathematical Society, 2009), xi-xii.

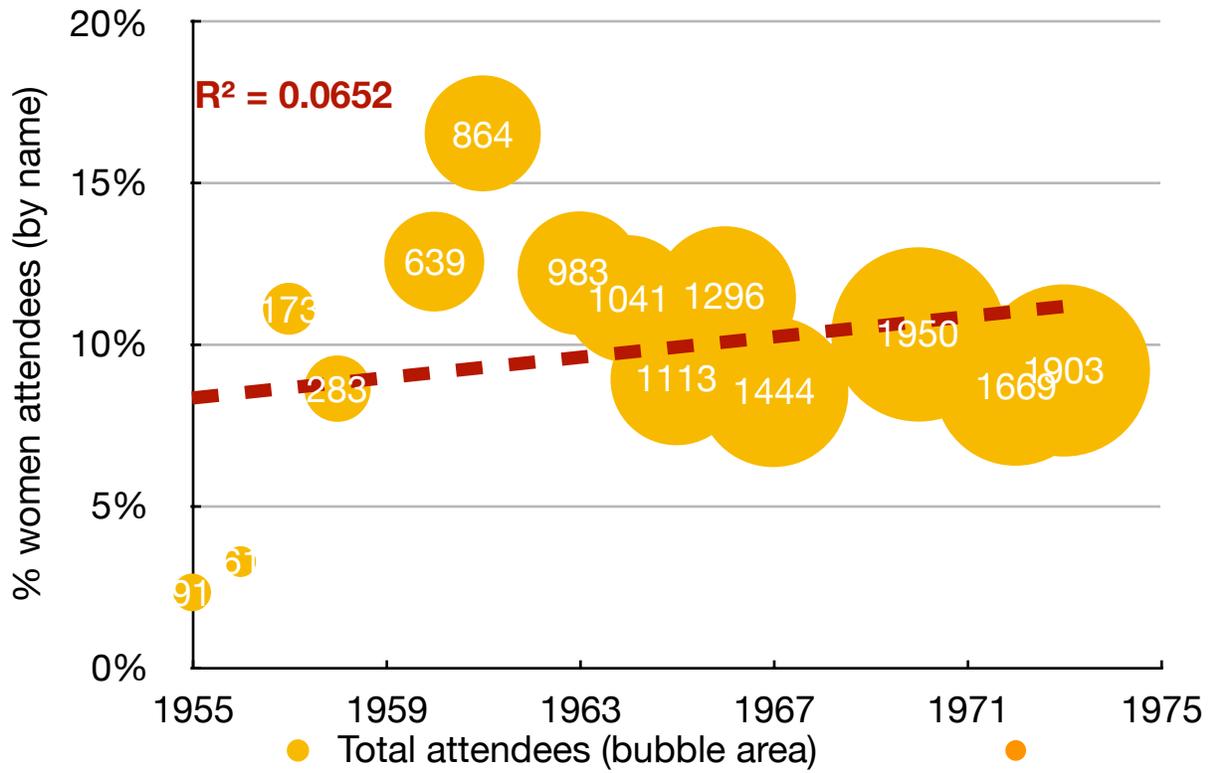

**Figure 3: Women's participation in SHARE (1955-1973)**

Attendance lists from SHARE show that it was dominated by men, especially in its first few years (see figure 3). Within two years of its founding, SHARE began a durable practice of organizing two large meetings a year. In its first months, however, there were organizational meetings in different parts of the country. I made a composite from the first three meetings in 1955; the very first such meeting had no women at all but then one woman attended each of the next two. The 1956 data point represents one regular meeting so its attendance appears anomalously lower. Participating organizations sent to SHARE their managers as well as their rank-and-file with attendees from such positions as manager, group supervisor, analyst, systems programmer, applications programmer, and catch-all "other." For one year (where this data was available), systems programmers were the largest single category followed by managers, "other," analysts, and group supervisors.[40]

---

[39] Atsushi Akera, "Voluntarism and the Fruits of Collaboration: The IBM User Group, Share," *Technology and Culture* 42 no. 4 (2001): 710-736 at https://doi.org/10.1353/tech.2001.0146 ; Jeffrey R. Yost, "The Origin and Early History of the Computer Security Software Products Industry," *IEEE Annals of the History of Computing* 37 no. 2 (2015): 46-58 at muse.jhu.edu/article/584412 . For background, see Roger W. Watt, "In support of users" [on SHARE] SIGUCCS '75 Proceedings of the 3rd annual ACM SIGUCCS conference on User services (1975): 58-61 at doi.org/10.1145/800115.803730

One measure of IBM's success in the computer marketplace was SHARE's large and increasing size. In the late 1960s when the Univac user group had around 300 members at its meetings, SHARE was 4 or 5 times larger and it grew to nearly 2000 attendees by 1970. IBM soundly dominated mainframe computing during these decades, and there is every reason to think that SHARE's membership was a representative sample of computer users across the country and (in time) around the industrial world. It seems unlikely that SHARE attendance data would be at odds with the wider field of computer programming (hypothetically) being fully 30 to 50 percent women in the 1950s and falling in the 1960s with (supposedly) "men's takeover of the field." Instead, the SHARE data supports quite the opposite. After the first few years, women consistently made up 8 to 16 percent of SHARE attendees with a rough "trend line" increasing from 9 to 12 percent. (With such a low $R^2$ value, it's unlikely there is any statistical significance.) The highest SHARE attendee level at 16.5 percent women is roughly half the hypothesized 30 percent.

SHARE's continual growth provoked logistical challenges. Its meetings became immensely complicated to organize, and its semi-annual *Proceedings* volumes became larger and fatter. The professional staff in SHARE's Chicago headquarters expanded to keep pace. With its publication costs "skyrocketing" SHARE shifted in the mid-1970s from printing and mailing three thick paperback volumes after each (semi-annual) meeting to instead publishing just two volumes per meeting, with a physical "volume 1" profiling the talks and presentations deemed of general interest to SHARE members while "volume 2" became a catch-all repository for the rest, eventually totaling a whopping 15,000 two-column pages on microfiche.[41] The attendance records became unmanageable, too; March 1973 was the last meeting where first names are available for all attendees. The printed attendance records then permanently switched to initials-only, symbolizing a shift from a first-name-basis community to a larger and more impersonal society.

To extend a statistical view beyond 1973, we can examine the first-name listings of SHARE's officers (see figure 4). SHARE was run by around 20 volunteer officers until 1976, when in the middle of that year its officer corps more than tripled to 85. The organization had originally been organized around "projects" such as compilers and time-sharing and a few years earlier had already adopted a "divisional structure" with a small phalanx of "managers" responsible for various technical areas and managerial concerns. In 1976 the organization added legions of sub-managers for these evolving areas so that by the late 1980s there were 250 officer-managers responsible for the organization's six divisions: SHARE-wide activities, Applications Architecture & Data Systems, Communications, Graphics & Integrated Systems, Management, and Operating Systems Support. Possibly with an eye to making its officers and managers readily identifiable by rank-and-file members, SHARE published their complete first and last names and featured them prominently in the physical volume 1.

---

[40] SHARE *Proceedings* 41 (13-17 August 1973): 1:37 (tabulation of position). Of the total registration of 1714, the "positions" were tabulated as manager (429), group supervisor (193), analyst (205), systems programmer (518), applications programmer (73), operations (19), and "other" (277).

[41] For explicit discussion of publication costs, see SHARE *Proceedings* 42 (4-8 March 1974), 3:1671 (skyrocketing).

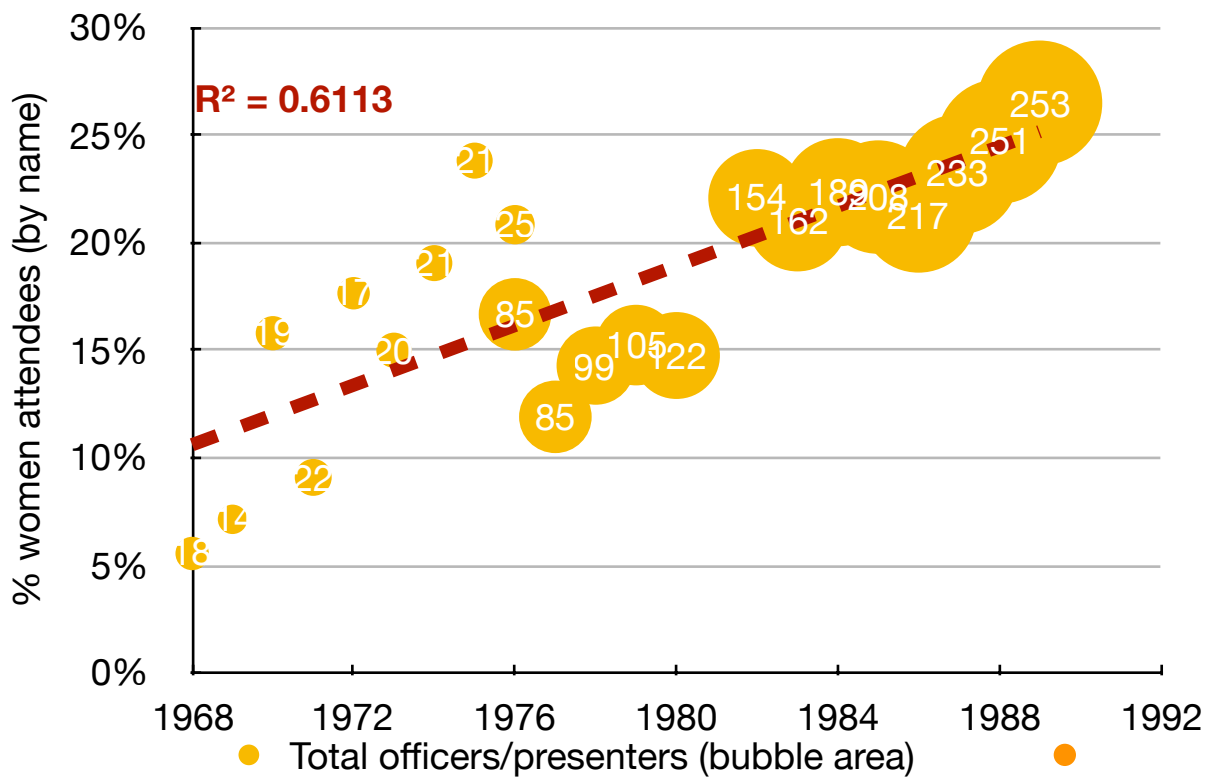

    Women's participation in SHARE leadership was substantial and growing throughout these years. Shirley F. Prutch from Martin-Marietta Data Systems became SHARE president in 1974, which led to some good-natured ribbing about her "coronation as the first Queen of SHARE" and, owing to her energetic leadership, the re-titling of SHARE as "Shirley Has Aided in Rejuvenation of Everyone."[42] Prutch was rising through the executive ranks in Martin-Marietta and in the mid-1980s became divisional vice president for systems integration and also

---

[42] SHARE *Proceedings* 43 (26-30 August 1974), volume 3:1622-26 on 1624 (queen) and 1625 (rejuvenation).

chair of a National Bureau of Standards panel on computer sciences and technology.[43] SHARE provided a valuable space for discussion about women flooding into computing.[44] Even before the 1976 expansion, women comprised generally 10 to 20 percent of SHARE's officers and managers and then rose to 26 percent by 1989.

The long-term growth of women in SHARE leadership—the trend line for officer-managers during 1968–1989 goes from roughly 10 percent to just over 25 percent with an $R^2$ value of 0.61 (moderate correlation)—is entirely consistent with the nationwide statistics on women's increasing proportion of computer science bachelor's degrees and women's increasing participation in the white-collar computing workforce. It is, of course, inconsistent with the notion of programming being fully 50 percent women or, especially, men staging some takeover of the field in the 1960s or 1970s.

**Data from Mark IV software user group**

What became the "Mark IV" software package had its origins in aerospace computing and the go-go years of the software-products industry in the 1960s. To tell a long story short, a 1962 spin-off from aerospace giant TRW (Thompson Ramo Wooldridge) called Informatics bought the software package's corporate owner and signed up its original designer who aimed a major new product, so-called Mark IV, at the brand-new IBM System/360. Informatics was led by Walter Bauer and Frank Wagner, and both had been aerospace company executives and user group leaders, respectively, in USE and SHARE. Mark IV enhanced the popular line of IBM 360s by offering to users structured forms that permitted "file creation, file maintenance, selection, extraction, processing, creating output files, sorting, and reporting," what we understand today as database management. Its sales really took off after IBM announced in 1969 that it would no longer "bundle" its software and hardware, neatly creating an open market that Mark IV stepped into squarely. In short order it smashed sales records right and left, eventually racking up an astounding $100 million in cumulative sales.[45]

Also in 1969 was the inaugural meeting for the Mark IV user group, sometimes labeled as the "Ivy League." Its female attendees included one "C. Ching" from Standard Oil of California. Later explicitly identified as Carol Ching, she was featured in an 1969 advertisement in the trade journal *Datamation* notable for positively valuing women as computer programmers.

---

[43] "Manufacturing," *Washington Post* (11 February 1985) at https://www.washingtonpost.com/archive/business/1985/02/11/manufacturing/41c52850-91d1-4e84-9d4c-f79aa8d27308/ ; "Executive Corner" *Computerworld* (9 January 1984): 81; Jane White, *A Few Good Women: Breaking the Barriers to Top Management* (Prentice Hall, 1992), 53-54.

[44] See SHARE *Proceedings* 43 (26-30 August 1974), 3:162.

[45] Richard Canning, "Data Management: Functions," *EDP Analyzer* 6 no. 1 (January 1968): 1-6, quote p. 2 (Mark IV description); Martin Campbell Kelly, *From Airline Reservations to Sonic the Hedgehog: A History of the Software Industry* (MIT Press 2003), p. 118.

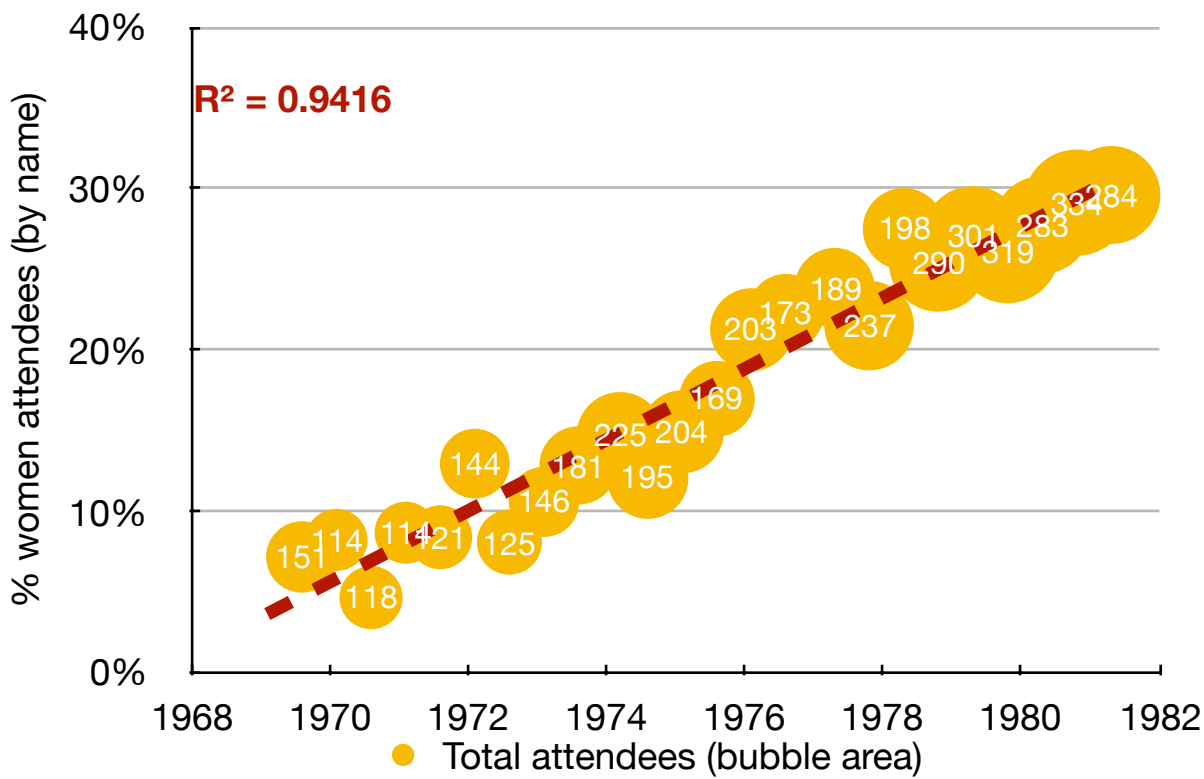

Figure 5: Women's participation in Mark IV (1969-1981)

Ampex was selling its magnetic tape, and the ad was formed around a personal image of her with the tag line "when programmer Carol Ching ignores our tape, we know we're doing our job." In an era when advertisements were sometime soaked in "Mad Men" style sexism, this matter-of-fact invoking of a female programmer was a sign that the culture of computing was changing.[46] And it was changing *not* to drive women out of the field, but rather recognizing that women were entering the field in increasing numbers. On the Mark IV data from 1969 to 1981 (see figure 5), women increased from somewhat under 10 percent to just over 30 percent of the user group attendees; and the $R^2$ value of 0.94 suggests reasonable significance for the trend line showing this increase. The 1981 figure of 32 percent is the first (and only) time in this dataset that women's participation topped 30 percent.

---

[46] William Vogel, "'The Spitting Image of a Woman Programmer': Changing Portrayals of Women in the American Computing Industry, 1958–1985," *IEEE Annals of the History of Computing* 39 no. 2 (2017): 49-64 on 52. An image of Carol Ching from 1970 appears in Dag Spicer, "Women in Computing through the Lens of *Datamation*," *CORE* (2016): 34-39 on 38. "Not until around 1970 does any explicit discussion of sexism or the need to examine and redefine gender assumptions appear in the data processing literature," notes Thomas Haigh in "Masculinity and the Machine Man: Gender in the History of Data Processing," in Misa, *Gender Codes*, 51-71, quote p. 63.

**Were women hidden somewhere?**

Given this new data on women in the computing workforce, one of two things must be true: either there are thousands of women somehow "missing" from this data set—or we must revise the common (but incorrect) image of women's numerical dominance in early computing as well as the (also mistaken) "takeover by men" of the field in the 1960s or 1970s. I believe it's the commonplace "memes," discussed in this chapter's introduction, that need revising. All the same, let's examine some possible weaknesses in the dataset.

I approached this data originally thinking that women *might* prefer to be known by their initials rather than by their gender-identifying first name. Several prominent computing women were widely known by traditional men's names, such as Stephanie 'Steve' Shirley and Elizabeth 'Jake' Feinler, who, respectively, founded an early woman-dominated software company (in 1962) and directed the Arpanet–Internet's Network Information Systems Center that created the top-level domain names such as .edu, .gov, .org, and .com.[47] Perhaps women preferred to be known by their gender-ambiguous initials and family names? With more than 15,000 names from SHARE alone, we have some data to consider.

There is little evidence that women in this dataset preferentially used initials or otherwise disguised their given first names. After "resolving" hundreds of initial-only names, it dawned on me that the balance of (resolved) women's and men's names were in proportion to the underlying balance of women and men. Where women were (say) 10 percent, it was roughly one in 10 initials-only names that could be identified as a woman; and where women were larger or smaller in the sample population, the pattern was roughly the same. Indeed, in successive years, the same person might be listed as C. Ching or E.A.S. Clark in one year and as Carol Ching and Anne Clark in the next. I can detect in this data no overarching "preference" expressed by women to use, or not use, initials for whatever reason. In name lists from the 1950s typescript was common and by the 1960s computer print-outs then laser-printed sheets were the chosen means. At a certain moment when the registration lists became truly immense, as with SHARE's nearly two thousand attendees, the easily formatted "initial-only" names might have looked cleaner or neater to the conference organizers. Either way, the use of initials does not seem a mechanism to hide women.

Another line of evidence suggesting that women were neither disproportionately hidden, nor for that matter revealed, by use of 'initials only' comes from closer examination of the SHARE rosters. In his analysis, Will Vogel observed that the proportion of women in SHARE meetings stayed consistent even when successive meetings varied widely in the use of initials-only attendance lists. For instance, for three years during 1958–61 the prevalence of initials-only in the registration lists nearly doubled from 24 to 47 then fell back to 22 percent, while the proportion of women in the gender-identified sample grew steadily from 9 to 15 percent (with an intermediate value of 13 percent when nearly half the meeting roster was initials-only). Even more dramatically, in four sample years during 1966–72, the prevalence of initials-only was as high as 48 percent and as low as 0.6 percent (in 1970), while the proportion of women in the gender-identified sample was steady around 8 percent—with actual year-by-year numbers of 8.5, 7.5, 8.6, 8.4 percent.[48] (See the two low-ball data points in the scatterplot above.) Surely if, hypothetically, hundreds of women were hidden behind initials-only names, they would have been revealed in 1970.

---

[47] Elizabeth Feinler, "Host Tables, Top-Level Domain Names, and the Origin of Dot Com," *IEEE Annals of the History of Computing* 33 no. 3 (2011): 74-79.

**Concluding thoughts**

Two conclusions seem reasonable based on the data presented in this chapter. First, the dataset on user groups is consistent with the 1970 census tabulation of women as 22.5 percent of the computing workforce. It's not surprising that SHARE, the largest such user group, is reasonably close to the Census's estimate (which recall is itself a *sample*). Women in the Mark IV user group passed 22.5 percent in the mid-1970s and reached just over 30 percent by the early 1980s, near the peak of women's share of employment in the computing industry. Still, it must be allowed that the user-group data samples might undercount women in the wider computing workforce since it's possible that more men than women from the membership organizations actually attended the user-group meetings (see tabulation of "positions" discussed above). This is one possible source of systemic bias that is not easily resolved, in the absence before 1970 of comprehensive firm-level or nation-wide data on computing women. Second, across the graphs in this chapter, women were an *increasing* portion of the computing workforce beginning in the 1960s and continuing through the late 1980s. There is no evidence from this data that men were staging a takeover of computer programming in the 1950s, 1960s, or 1970s.

This chapter provides new data on the computing workforce suggesting that women were a prominent but relatively small proportion of the skilled computing workforce in the 1950s and 1960s. In these years, I think it is more likely that women were around 15 percent of the skilled computing workforce than the 30 to 50 percent that is now widely accepted. A figure of roughly 15 percent women is consistent with trade literature and professional publications, images in technical and popular media, dozens of archival photographs, and periodic salary surveys of the computing workforce.[49] The data in this chapter strongly supports that women were an *increasing* proportion of the skilled computing workforce beginning in the mid-1950s through to the peak in the mid-1980s. The data is entirely inconsistent with any suggestion of a male "takeover" of computing sometime in these decades. To repeat the obvious, women were flooding into computing during these years—not being chased out. I have also suggested why the inaccurate but possibly comforting image of the male takeover, and its connection to a "linear storyline," has taken hold of our imagination.

---

[48] William Vogel "Women Programmers in the 1950s and 1960s: SHARE Statistics," *CBI Newsletter* 37 no. 2 (Fall 2015): 25-27. Vogel's numbers are slightly different from the ones I report here; he and I "found" the same amount of women, but I tended to "resolve" more initial-only and gender-ambiguous names (using the SSA data). I also used standard statistical sampling (confidence level 95%, p 0.5, error 0.05) when the meetings grew larger than 800 (from 1961), while he tallied all SHARE attendees (up to 1,950 names).

    The chapter has one longer-term implication for understanding gender bias in computing today. If, hypothetically, men staged a take-over in the 1950s or 1960s with the aim of raising the status of the computing profession by ridding it of lower-status women—such "feminization" is discussed in the sociological literature[50]—the clear implication is that gender bias and sexism was "baked into" the computing profession during the years that it was forming. Ridding a profession of such core values might be difficult indeed. But this data supports a different viewpoint entirely: it suggests that gender bias is not a foundational or core value of computing professionals, since computing as a profession took form during the years when women were flooding into computer science and the skilled computing workforce. The problem of gender bias in computing today is not to be located in 1960s sexism but the more recent cultural and social dynamics of the mid-1980s. Further research is needed on why women entered the computing profession and skilled workforce, what their experiences were during those years of expanding educational and workforce opportunities, and how the more complex subsequent history bears on the current problem of gender bias in computing.

---

[49] In its 1960 salary survey the trade journal *Business Automation* found "Less than 15 percent of the [computer] programmers reported were women," quoted in Thomas Haigh, "Masculinity and the Machine Man: Gender in the History of Data Processing," in Misa, *Gender Codes*, 51-71, p. 54.

[50] See Sharon Hartman Strom, "'Machines Instead of Clerks': Technology and the Feminization of Bookkeeping, 1910-1950," in Heidi Hartmann, ed., *Computer Chips and Paper Clips: Technology and Women's Employment*, Volume II: *Case Studies and Policy Perspectives* (Washington DC: National Academy Press, 1987), 63-97 at www.nap.edu/read/951/chapter/4 ; Rosemary Wright and Jerry A. Jacobs, "Male Flight from Computer Work: A New Look at Occupational Resegregation and Ghettoization," *American Sociological Review* 59 no. 4 (1994): 511-536 at doi.org/10.2307/2095929 ; and Marie Hicks, "Meritocracy and Feminization in Conflict: Computerization in the British Government," in Misa, *Gender Codes*, 95-114.